\documentclass[prl,aps,notitlepage,tightenlines,twocolumn]{revtex4-1}
\usepackage{amsmath}
\usepackage{amssymb}
\usepackage{bm}
\usepackage{graphicx}
\usepackage{wrapfig}
\usepackage{natbib}
\usepackage[colorlinks=true,urlcolor=blue,bookmarks=true,citecolor=magenta,breaklinks=true]{hyperref}
\usepackage{enumerate}
\usepackage{xcolor}

\newcommand{\brho}{\bm{\rho}} 
\newcommand{\rvec}{\mathbf{r}} 

\begin{document}
\title{The biexciton puzzle}
\author{David K. Zhang, Daniel W. Kidd, and K\'alm\'an Varga}
\affiliation{Department of Physics and Astronomy, Vanderbilt University, Nashville, Tennessee 37235, USA}

\begin{abstract}
	The Stochastic Variational Method (SVM) is used to show that the effective mass model 
	correctly estimates the binding energies of excitons and trions, but fails to predict 
	the experimental binding energy of the biexciton. Using high-accuracy variational calculations, 
	it is demonstrated that the biexciton binding energy in
	transition metal dichalcogenides is smaller than the trion
	binding energy, contradicting experimental findings. It is
	also shown that an excited state of the biexciton is in very
	good agreement with experimental data. This excited state
	corresponds to an hole  attached to a negative trion 
	and may be a possible resolution of the discrepancy between theory and experiment.
\end{abstract}

\maketitle

Since the isolation of graphene, much attention has been directed toward two-dimensional (2D) materials and their extraordinary potential within nanoelectronics and photoelectronics \cite{kis1,kis2,dark_exciton,trion_MoS2,charged_exc,nl501988y,pseudospin,optical_gen,PhysRevLett.113.026803}. Of specific interest in recent research has been the development of a graphene-like 2D material featuring a sizable band gap, allowing it to serve as a 2D semiconductor. In particular, transition metal dichalcogenides (TMDs) have been investigated as popular candidates.
An observed consequence of reduced dimensionality and weak dielectric screening in such materials is a strong electrostatic interaction allowing the formation of bound electron-hole pairs (excitons) with binding energies on the scale of several hundred meV in TMDs such as $\text{WSe}_2$ \cite{PhysRevLett.113.026803}. This behavior is not observed within 3D bulk counterparts and is thus notably unique to few-layer materials. Charged excitons (trions) have also been observed with surprisingly large binding energies in $\text{MoS}_2$ (20 meV) \cite{Mak2013}, $\text{WS}_2$ (45 meV) \cite{Zhu12082014}, and $\text{WSe}_2$ (30 meV) \cite{PhysRevB.90.075413}. Such large exciton and trion binding energies suggest the existence of bound exciton pairs (biexcitons) in monolayer TMDs, which have indeed been experimentally found in $\text{WSe}_2$ (52 meV) \cite{You2015} and $\text{WS}_2$ (65 meV) \cite{1507.01342}.

There is a substantial need to understand these excitonic structures in 2D materials in order to characterize their electrical and optical responses and fully assess their potential functionality. Effective-mass models are often used to model electron-hole systems \cite{PhysRevLett.1.450,PhysRevB.59.5652,PhysRevB.61.13873} and have been widely applied to excitonic systems in 2D materials \cite{Keldysh,PhysRevB.91.245421,PhysRevB.88.045318,1505.03910,PhysRevLett.114.107401}. These models have been successful in the calculation of binding energies and geometries (extended wavefunctions) of excitons and trions in TMDs.

In this work, we show that while the effective mass model correctly estimates the binding energies of excitons and trions, \textit{it fails to predict the binding energy of the biexciton}. Using high-accuracy variational calculations, we demonstrate that the biexciton binding energy predicted in TMDs by the effective mass model is smaller than the corresponding trion binding energy, contradicting the aforementioned experimental findings. However, we also show that the binding energy of an excited state of the biexciton closely agrees with experimental data. This state corresponds to an electron attached to a positive trion and may resolve the discrepancy between theory and experiment.

The computational technique employed in this paper is the Stochastic Variational Method  
applied to exciton ($\text{X}=eh$), trion ($\text{X}^-=eeh$), and biexciton 
($\text{X}_2=eehh$) systems using the explicitly correlated Gaussian (ECG) basis. This basis is known to produce high-accuracy binding energies (8-10 decimal digits) when applied to similar few-particle systems, including $\text{H}_2$ \cite{Bubin200912}, $\text{H}_2^+$, and the positronium molecule $\text{Ps}_2$ \cite{PhysRevA.47.3671,PhysRevA.48.1903}.

The nonlinear parameters of these explicitly correlated Gaussians are optimized using stochastic variation, a procedure which uses random trial and error to iteratively improve the quality of the ECG-basis representation of the desired wavefunction. This method is known to be well-suited to the description of both ground and excited states of excitonic structures with up to five particles, such as positively charged biexcitons $\text{X}_2^+$ ($eehhh$) \cite{PhysRevB.59.5652,PhysRevB.78.235307}. We refer the reader to Ref.\@ \cite{appliedECG} for a through review of the applications of the ECG basis in various problems.

The nonrelativistic Hamiltonian of an excitonic $N$-particle system is given by
\begin{equation} \label{Hamiltonian}
	H = -\frac{1}{2} \sum_{i=1}^N \frac{1}{m_i} \nabla^2_i
	+ \sum_{i<j}^N \frac{q_i q_j}{r_s}
	V \left( \frac{r_{ij}}{r_s} \right)
\end{equation}
where $\rvec_i$, $m_i$, and $q_i$ are respectively the 2D position vector,
effective mass, and charge of the $i$th particle, $r_{ij} =
|\rvec_i - \rvec_j|$,
and $r_s$ is the screening length of the medium. The 2D screened electrostatic
interaction potential is then given by
\begin{equation} \label{Screened2DPotential}
	V(r) = \frac{\pi}{2} [H_0(r) - Y_0(r)]
\end{equation}
where $H_0$ and $Y_0$ are the Struve function and Bessel function of the second
kind, respectively. This potential differs substantially from the usual $1/r$
Coulomb potential, exhibiting nonlocal macroscopic screening which arises in 2D
systems \cite{rubio}. For small distances ($r \to 0$) the potential exhibits
logarithmic divergence, while for large distances ($r \to \infty$) it asymptotically
falls off as a $1/r$ Coulomb potential. The 2D layer
polarizability $\chi_{2D}$ determines the length scale $r_s = 2\pi\chi_{2D}$ that
separates these two regimes.
\par\indent
As a variational trial function, we adopt a 2D form of the correlated Gaussians
\cite{svmbook,appliedECG}
\begin{equation}
\Phi_M(\rvec) = \mathcal{A} \left\{
	\left( \prod_{i=1}^N \xi_{m_i}(\brho_i) \right)	
	\exp\left( -\frac{1}{2} \sum_{i,j=1}^N A_{ij} \brho_i \cdot \brho_j \right)
\right\},
\end{equation}
where
\begin{equation}
\xi_m(\bm{\rho}) = (\rho_x + i\rho_y)^m.
\end{equation}
Here, $\brho_i$ denotes the $i$th relative (Jacobi) coordinate of the system,
$m_i$ denotes the magnetic quantum number of the $i$th particle,
and $M = m_1 + m_2 + \cdots + m_N$. This function is
coupled with the spin function $\chi_{SM_S}$ to form the trial
function. 

The material-dependent parameters (effective masses and screening lengths)
adopted in this paper are those given in Ref.\@ \cite{PhysRevB.88.045318}. These parameters
are based on ab initio calculations and produce exciton binding energies in
close agreement with Bethe-Salpeter calculations. For simplicity,
we fix the electron-hole mass ratio $\sigma = m_e / m_h$ to unity,
and consider only negative trions ($eeh$) in our calculations.
(No loss of generality is incurred, as when $\sigma = 1$,
charge conjugation symmetry guarantees that the
energies produced are equivalent to those of
positive trions.) Indeed, experimental observations find that the binding energies
of positive and negative trions are nearly equal \cite{charged_exc},
validating our choice of $\sigma = 1$ as a good 
approximation. Even so, we point out that slight variations in $\sigma$
do not affect our qualitative results.

The calculated and experimental binding energies of excitons are
reported in Table I. Note that the experimental binding energies are
in the 500 meV range
\cite{dark_exciton,optical_gen,nl501988y,PhysRevLett.113.026803} and the
parameters given in Ref.\@ \cite{PhysRevB.88.045318} give good overall
agreement with experiment. More accurate agreement is not pursued,
partly due to the presence of uncertainties in the experiments
(e.g.substrate dependence),
and partly because the binding energies of the trions and biexcitons are
not very sensitive to the exciton binding energy. The present SVM approach reproduces
the energies of Ref.\@ \cite{PhysRevB.88.045318} for excitons and improves
the results of Ref.\@ \cite{PhysRevB.88.045318} by a few meV for trions.
This is reasonable because we use a variational ansatz with $100$
basis functions, while in Ref.\@ \cite{PhysRevB.88.045318} only a
single trial function is used. 

The experimental observation of biexcitons in monolayer WSe$_2$ reported in Ref.\@
\cite{You2015} is accompanied by a variational calculation estimating 
the binding energy to be 37 meV with a one-term variational trial function.
This is lower than the measured value of 52 meV reported in the same paper.
The authors expect that a more accurate approach would reconcile theory and
experiment, but our calculation shows that this is not the case. A larger
basis provides improved estimates of both exciton and biexciton ground-state
energies, giving a biexciton binding energy even smaller than the single
term prediction in Ref.\@ \cite{You2015} (see Table I).

The underbinding of the biexciton is not unique to WSe$_2$. 
We consistently observe that for typical screening lengths occurring in
MoS$_2$, MoSe$_2$, WS$_2$, and WSe$_2$, the biexciton binding energy
is always significantly smaller than the trion binding energy (see Table I).
This stands in contrast to experiment, where trion binding energies are
usually found in the 20-30 meV range, and biexciton binding energies are
found between 50 and 70 meV. A general trend appearing in our calculations
is that above a certain screening length ($r_s>5$ a.u.) the
biexciton is less strongly bound than the trion, while for smaller screening lengths
($r_s<5$ a.u.) the biexciton becomes more strongly bound.
However, even this adjustment in screening length cannot be made to agree
with experiment, since in this range the exciton binding energy
differs significantly from the observed value. (The exciton binding energy
increases to 2.5 eV, while the trion binding energy becomes 250 meV).
Varying the electron-hole mass ratio $\sigma$ would slightly change the
results but again would not help to resolve the disagreement.

In addition to its ground state, the biexciton also has three bound excited
states \cite{PhysRevA.58.1918,PhysRevB.59.5652,Suzuki200067}.
Two excited states exist with $L=0$ and positive parity, and one exists with $L=1$ and negative
parity \cite{PhysRevA.58.1918}. One of the $L=0$ excited states is
bound due to charge inversion symmetry, but this is only valid in
the case of equal electron and hole masses. Thus, we have not considered this state 
in our calculations. The $L=0$ excited state is bound with respect to the
$\text{X}(1S)$+ $\text{X}(2S)$ threshold
(where $\text{X}(nL)$ denotes the $nth$ exciton state with orbital momentum
$L$), and the $L=1$ state is  bound with respect to the 
$\text{X}(1S)$+ $\text{X}(2P)$ channel.
(Note that while the usual $1/r$ Coulomb potential gives equal energies
for the $2S$ and $2P$ exciton states, this degeneracy is broken
by the screened 2D potential under present consideration.)
These excited states cannot autodissociate to the $\text{X}(1S)$+$\text{X}(1S)$
threshold, as symmetry considerations
\cite{PhysRevA.58.1918,Suzuki200067,PhysRevLett.92.043401}
force this channel to be closed. 

\begin{figure}[htp]
  \includegraphics[width=.33\textwidth,clip=true]{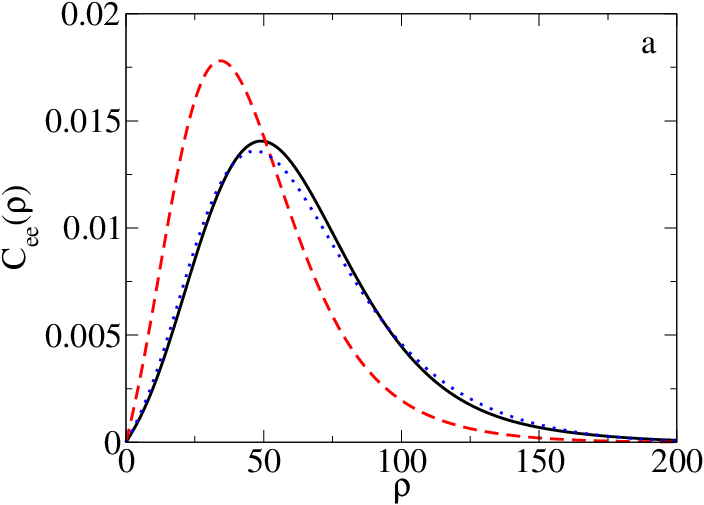}
  \includegraphics[width=.33\textwidth,clip=true]{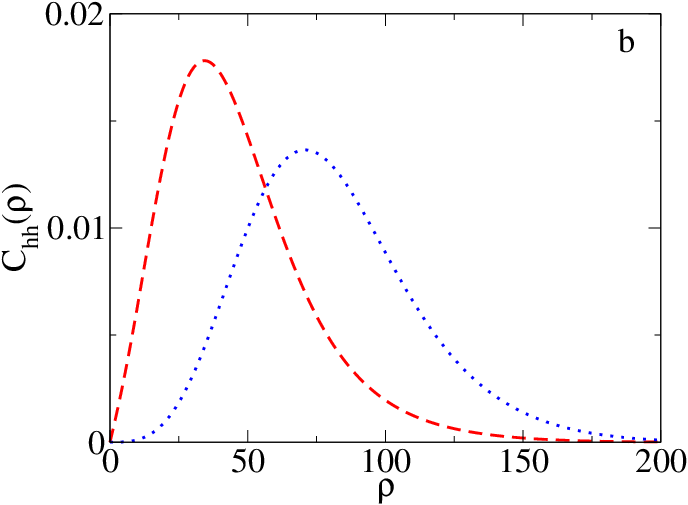}
  \includegraphics[width=.33\textwidth,clip=true]{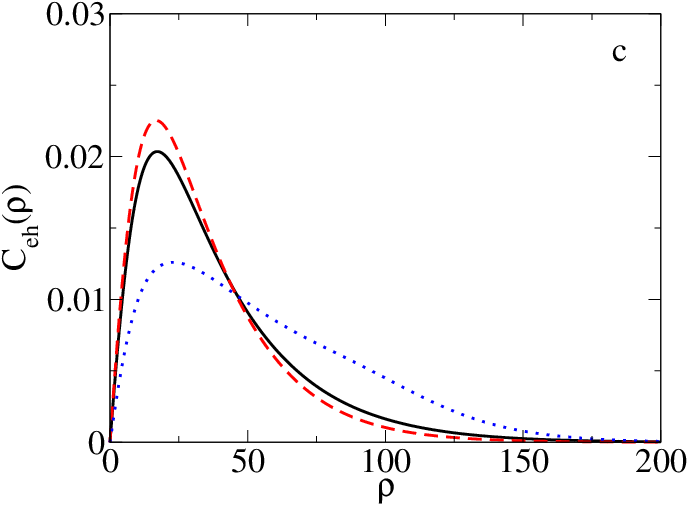}
\caption{(Color online) Electron-electron (a) hole-hole (b),
electron-hole (c) correlation functions for the $\text{WSe}_2$ trion (solid line)
ground-state biexciton (dashed line), and $L=0$ excited biexciton (dotted line). Atomic units
used.}
    \label{fig:pd}
\end{figure}

Our calculations show that the energy of the $L=1$ excited state is lower
than that of the $L=0$ excited state, while simultaneously the
threshold of the $L=1$ state is higher than the threshold for the
$L=0$ state. This results in binding energies of order of 250-450 meV (see
Table I), much larger than experimental results. Moreover, this
$L=1$ state is much less likely to be formed than the $L=0$ state. 

\begin{figure}[htp]
  \includegraphics[width=.33\textwidth,clip=true]{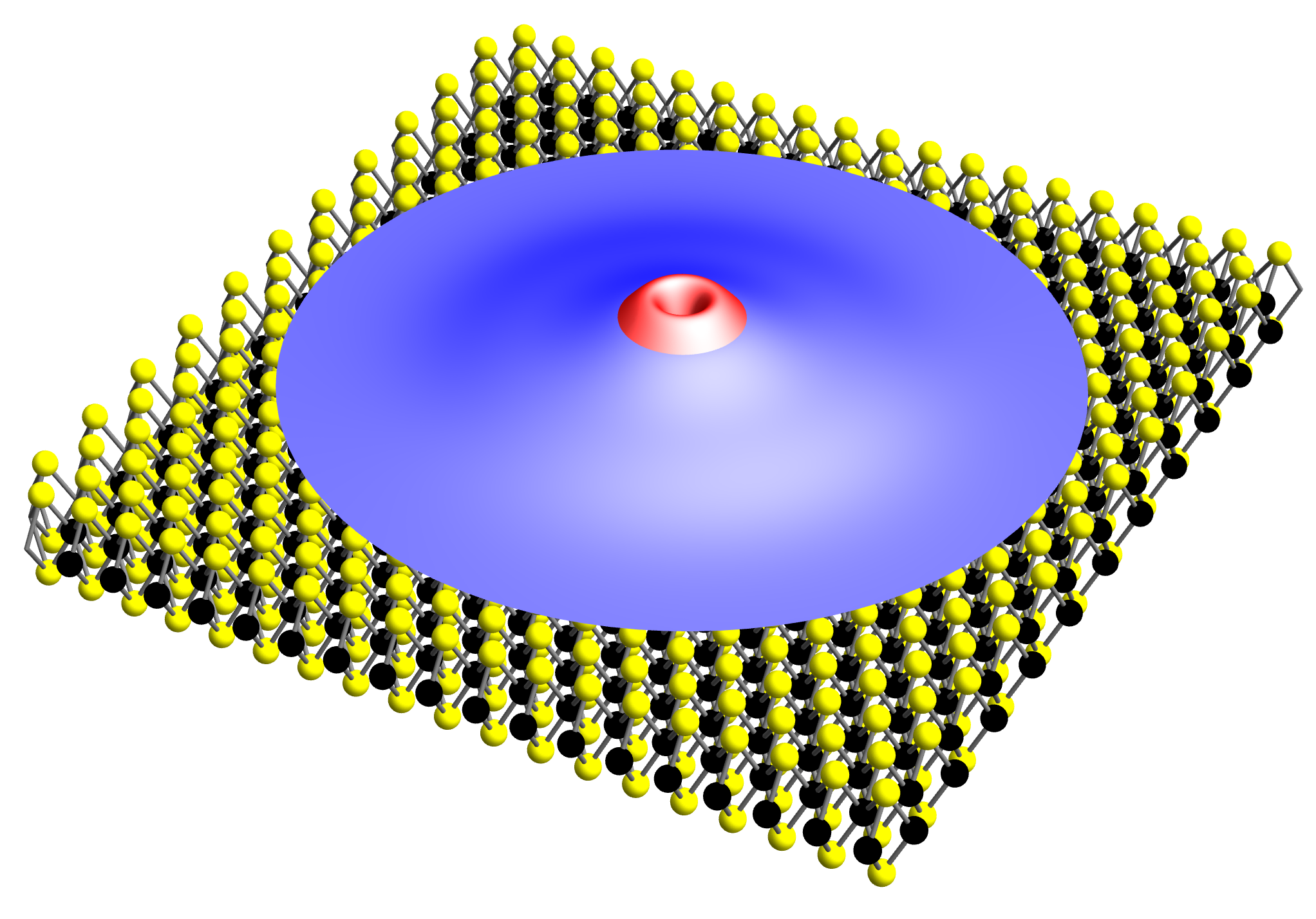}
\caption{(Color online) Schematic view of the excited biexciton as a
system of a trion core (red) and a long tail representing the hole (blue).}
\end{figure}

\begin{table*}
\centering
\caption{Comparison of experimental and theoretical TMD excitonic
binding energies (meV).}
\begin{tabular}{|l|l|p{2.2cm}|p{2.2cm}|p{2.2cm}|p{2.2cm}|}\hline
& System               & $\text{MoS}_2$ & $\text{MoSe}_2$ & $\text{WS}_2$ & $\text{WSe}_2$ \\ \hline
 Experiment& $\text{X}$       & 500\cite{nl403742}, 570\cite{exciton_mos2}
 & 550\cite{exciton_mose2}             &
 320\cite{PhysRevLett.113.076802}, 700\cite{dark_exciton}       &
 370\cite{PhysRevLett.113.026803}            \\
 Theory& $\text{X}$           & 555            & 480             & 523       & 470          \\ \hline
 Experiment& $\text{X}^-$     & 18$\pm$1.5\cite{Mak2013} &
 30\cite{charged_exc,PhysRevLett.112.216804} &
 30\cite{1507.01342}, 45\cite{Zhu12082014} & 30\cite{optical_gen,PhysRevB.90.075413}             \\
 Theory& $\text{X}^-$         & 34               & 28              & 34
 & 30             \\ \hline
 Experiment& $\text{X}_2$     & 70\cite{nl403742} & --              & 65\cite{1507.01342}           & 52\cite{You2015}             \\
 Theory& $\text{X}_2$         & 22             & 18              & 24           & 20             \\
 Theory& $\text{X}_2^*\ (L=0)$ & 69             & 58              & 67
 & 59             \\
 Theory& $\text{X}_2^*\ (L=1)$ & 460            & 430             & 360
 & 240          \\ \hline
\end{tabular}
\end{table*}

To investigate the structure of the excited state, we studied the pair correlation function
\begin{equation} \label{corr_func_pair}
	C_{pq}(\rvec) = \frac{2}{N(N-1)} \left\langle \Psi \left| \sum_{i<j}^N \delta(\rvec_i-\rvec_j-\rvec) \right| \Psi \right\rangle,
\end{equation}
where $p$ and $q$ stand for electrons or holes, and the sum is taken only over
corresponding pairs. Fig.\@ 1(a) shows the electron-electron ($ee$)
correlation functions. Those of X$^-$ and X$_2$ are quite different. By
adding a hole to X$^-$ and thus forming a singlet state with the hole  of the
trion, the resulting X$_2$  becomes more compact than the trion due to the
extra attraction. The $ee$ correlation function of the 
X$^-$ and X$_2^*(L=0)$ are, however, almost identical, indicating that
the hole (which is in this case coupled as a triplet with the hole of
the trion) is situated somewhere far away from the trion and thus does
not disturb the structure of the trion.

Fig.\@ 1(b) shows the hole-hole ($hh$) correlation functions. These
are different for the ground and excited state of the biexciton, as
expected from the previous discussion. The electron-hole ($eh$) correlation
functions, shown in Fig.\@ 1(c), are very similar in X$^-$ and X$_2$;
the electron and hole take up the most energetically favorable positions
in the trion, and because there is no symmetry restriction,
the same occurs in the ground state biexciton.
In X$_2^*(L=0)$, however, the spin triplet nature of the hole wavefunction 
restricts the available spatial positions. The peak of the 
X$_2^*(L=0)$ $eh$ correlation function is at the same position but
roughly half the amplitude of that of X$^-$ and X$_2$, indicating
that one of the holes exists in a trion-like structure. The tail of the
X$_2^*(L=0)$ correlation function extends far beyond the other
correlation functions, showing that a hole exists somewhere outside of
X$^-$ and corroborating the X$^-+h$ structure shown in Fig.\@ 2. 
These extended correlation functions also suggest that simple one-term wavefunctions
are unlikely to provide accurate descriptions of these structures.

In summary, by using high-accuracy variational calculations, we have shown that 
in models aiding experiments, the binding energy of biexcitons
in transition metal dichalcogenides is less than half of the
experimental value. An excited biexciton, however, is
a possible candidate for the experimentally observed state.
Investigations of other effects, e.g. surface-bound or
defect-bound excitons or excitonic complexes are necessary before the 
experimental data can be unambiguously assigned to a new biexciton
formation.

%


\end{document}